\documentclass[english]{revtex4-2}
\usepackage[T1]{fontenc}
\usepackage[latin1]{inputenc}
\usepackage{color}
\usepackage{bm}
\usepackage{amstext}
\usepackage{amssymb}
\usepackage{graphicx}
\usepackage{esint}

\makeatletter

\usepackage{babel}

\makeatother

\usepackage{babel}
\begin{document}
\title{External charge perturbation in a flowing plasma and electrostatic
turbulence}
\author{Mridusmita Das\footnote{mridusmitadas1993@gmail.com} and Madhurjya P. Bora}
\address{Physics Department, Gauhati University, Guwahati 781014, India.}
\begin{abstract}
In this work, an 1D electrostatic \emph{hybrid}-Particle-in-Cell-Monte-Carlo-Collision
(\emph{h-}PIC-MCC) code  is used to study
the response of a plasma to a moving, external, charged perturbation
(debris). We show that the so-called pinned solitons can form \emph{only}
under certain specific conditions through a turbulent regime of the
ion-ion counter-streaming electrostatic instability (IICSI). In fact,
the pinned solitons are manifestation of the ion phase-space vortices
formed around the debris. The simulation shows that the pinned solitons
can form \emph{only} when the debris charge density exceeds a certain
value causing the counter-streaming ion velocity to exceed a critical
velocity, pushing the instability to a turbulent regime. The effect
of debris velocity is also essential for the appearance of pinned
soliton as when the debris velocity increases, it causes the widening
of the phase space vortices causing well-separated pinned solitons,
which merge to form one single soliton when debris velocity reduces
to zero. In the opposite extreme, when debris velocity becomes highly
supersonic, the vortices are widened up to a limit causing the pinned
solitons to disappear altogether. We further show the existence of
a Kolmogorov-type energy cascade scaling  for this electrostatic
turbulence. 
\end{abstract}
\maketitle

\section{Introduction}

Despite being studied actively and widely since the days of Irving
Langmuir (early 1920s), certain fundamental issues in plasma physics
continue to enjoy the attention of the scientific community and prove
their importance toward understanding the complex behavior of plasmas.
In recent years, there has been a considerable interest in studying
the response of a flowing plasma to an externally embedded charged
perturbation (so-called debris), both theoretically \citep{fkdvleo1,fore-wake,mag,sarkar}
and experimentally \citep{pinned,precursor,size-shape}. One of the
reasons for interest in this kind of problems is, in principle, exploring
the possibility of detection of space debris in low-earth orbits (LEO)
\citep{fkdvleo1}. Historically, the effect of the movement of a charged
particle through a plasma was studied quite a long time ago in 1955
\citep{passage-old}. Subsequently, several authors have studied the
formation of wakes, generated by such movements in $e$-$i$ as well
as in complex plasmas \citep{wake1,wake2,wake3}. The formation of
wakes with reference to laser-plasma interaction was also studied
by Malka and several others \citep{laser}.

A localized charge perturbation in a plasma can primarily occur in
two different ways -- due to accumulation of charges on the surface
of an external body such as debris which is embedded in the plasma
and due to the formation of polarized structures as a result of self-consistent
nonlinear interactions within the plasma itself \citep{debye-scale}.
In both the cases however, the charge perturbation, due to its localized
nature, influences the plasma particles (both ions and electrons)
in the neighborhood which can lead to the formation of nonlinear structures
with interesting dynamics. One can find a number of theoretical works,
which are devoted to the investigation of the formation of nonlinear
structures due to external charge perturbations (debris) \citep{fkdvleo1,sarkar,fore-wake}
theoretically as well as through molecular dynamics simulations \citep{fore-wake,mag}.
Several authors have also studied the effect of size and shape of
these charged debris experimentally in the dust-acoustic regime using
a complex plasma device \citep{size-shape,pinned}.

On the other hand, the self-organization of nonlinear structures in
plasmas can give rise to Debye-scale polarized structure which can
then act as a site of localized perturbation \citep{debye-scale,murchana}.
With sufficient strength, these Debye-scale structures can give rise
to streaming instabilities. An excellent review of these types of
structures and their interactions can be found in a paper by Schamel
\citep{hans}. In a recent work, Wang et al. \citep{debye-scale}
have carried out a statistical analysis of several bipolar electrostatic
structures in the bow shock regions of Earth and argued that these
bipolar structures \citep{vasco} are ion phase-space holes produced
by the two-stream instability triggered by the incoming and reflected
ions in the shock transition region. These structures were detected
by the \emph{Magnetospheric Multiscale }(MMS) spacecraft \citep{mms}.

Toward this, we in this work, explore the response of a plasma to
a moving external charge perturbation through particle-in-cell (PIC)
simulation. Particularly, we show that \emph{only} a certain kind
of charge perturbation leads to the formation of the so-called pinned
solitons, which can be the manifestation of electrostatic turbulence,
driven by an ion-ion counter-streaming instability (IICSI). The simulation
itself is being carried out with our well-tested \emph{hybrid-}PIC-MCC
code \citep{suniti1,suniti2,suniti3}. However, we should be cautious
to mention that in the \emph{h-}PIC-MCC code, the Monte-Carlo collision
algorithm is used only for collision of charged particles with dust
particles, required for the purpose of charging of the dust particles.
For all other cases, the simulation remains collision-less. Historically,
two-stream instability by counter-streaming particles is quite well
understood in the framework of kinetic theory \citep{history1,history2,history3}.
They are also studied in the context of particle beam ramming through
a plasma, both in classical and relativistic situations \citep{two-stream-beam,relativistic}.
However, there are a couple of fine points where we would like to
draw the attention of the reader, especially the nonlinear saturation
of the instability, where phase-space holes can sustain within a well-developed
turbulence scenario \citep{two-stream}. In fact, in one of the works
\citep{turb}, it has been argued that ion-ion counter-streaming turbulence
\emph{cannot} possibly lead to the formation of electrostatic shocks,
which also agrees with our simulation results.

In Section II, we review the basic theoretical framework for the formation
of these Debye-scale structures triggered by the presence of charged
debris \citep{sarkar,fkdvleo1}. In Section III, we present our theoretical
model (kinetic) for counter-streaming ion-ion instability and carry
out a linear stability analysis for this particular situation. In
Section IV, we present the results of our PIC simulation, which show
the formation of dissipation-less shock waves (DSW) and pinned solitons
for a charged external perturbation (debris). In Section V, we show
how a negatively charged external perturbation can lead to a turbulent
regime, only when the perturbation exceeds a certain threshold. Here,
we show that the pinned solitons are basically a manifestation of
the turbulent counter-streaming ion-ion instability. We also show
that, not surprisingly, the turbulence has a Kolmogorov-type energy
cascade scaling \citep{kol1}. In this section, we estimate the minimum
critical velocity required by the counter-streaming ion beams for
sustaining the instability and compare it with our theoretical estimate.
Toward the end, we have a short discussion on the effect of negatively
charged dust particles on pinned solitons with results closely agreeing
with other reported works \citep{sarkar}. Finally, in Section VI,
we give a brief summary of our findings and concluding remarks.

\section{Debye-scale structures in a flowing plasma}

In this section, we are going to briefly review the theoretical formalisms
for different nonlinear Debye-scale structures, namely pinned solitons
and DSWs in a flowing plasma. In order to investigate these nonlinear
phenomena, we consider a 1D, collision-less $e$-$i$ plasma with
an external charge perturbation (debris) of charge density $\rho_{{\rm deb}}$.
The debris charge density along with its shape and size are held constant
throughout the simulation. The equations are continuity and momentum
equations for ions, and Poisson's equation. The electrons can be considered
to be inertialess in the ion-acoustic (IA) timescale, 
\begin{eqnarray}
\frac{\partial n_{i}}{\partial t}+\frac{\partial}{\partial x}(n_{i}v_{i}) & = & 0,\label{eq:cont-1}\\
n_{i}\frac{dv_{i}}{dt} & = & -\sigma\frac{\partial n_{i}}{\partial x}-n_{i}\frac{\partial\phi}{\partial x},\\
\frac{\partial^{2}\phi}{\partial x^{2}} & = & n_{e}-n_{i}-\rho_{{\rm deb}},\quad n_{e}\sim e^{\phi},\label{eq:pois-1}
\end{eqnarray}
where electrons are assumed to be Boltzmannian. The above equations
are expressed in normalized forms where we have expressed the ion
pressure $p_{i}=n_{i}T_{i}$. The quantity $\sigma=T_{i}/T_{e}$ and
temperatures $T_{i,e}$ are expressed in energy units. The charge
density of the debris is denoted by $\rho_{{\rm deb}}\equiv\rho_{{\rm deb}}(x-v_{{\rm deb}}t)$,
which is moving with a velocity $v_{{\rm deb}}$. We note that $\rho_{{\rm deb}}\lessgtr0$
depending on the nature of the debris charge. This model predicts
the formation of pinned solitons \citep{fkdvleo1} as well as DSWs
\citep{sarkar} in the nonlinear regime under suitable conditions.
In the above equations, the densities are normalized by their equilibrium
values and electrostatic potential $\phi$ is normalized by $(T_{e}/e)$.
Length is normalized by electron Debye length and time is normalized
by $\omega_{pi}^{-1}$, the inverse of ion plasma frequency. All velocities
are normalized by the ion-sound speed $c_{s}=\sqrt{T_{e}/m_{i}}$.

We realize that depending on the nature and magnitude of $\rho_{{\rm deb}}$,
the response of the plasma can be quite different. While, for $\rho_{{\rm deb}}>0$,
we might see the formation of DSWs in the precursor region, pinned
solitons may form when $\rho_{{\rm deb}}<0$. Due to the presence
of a strong accumulation of positive charge $(\rho_{{\rm deb}}>0)$,
rapidly moving ions away from the debris site compresses the plasma
in the precursor region which leads to the formation of DSWs. For
sufficiently large $\rho_{{\rm deb}}<0$, IICSI may develop and results
in an electrostatic turbulence, causing phase-space vortices to form
which can effectively trap the ions spatially as well as temporally
at the site of the debris. This trapping of ions is what causes the
pinned solitons to form.

\subsection{Pinned solitons \& DSWs}

The formation of pinned solitons can be understood through the forced
Korteweg-de Vries (fKdV) equation, which can be obtained following
the well-established reductive perturbation analysis. To start with,
we expand the variables $g=(n_{i},v_{i},\phi,\rho_{{\rm deb}}\mathclose{)}$
with respect to an expansion parameter $\epsilon\,(\ll1)$ 
\begin{equation}
g\sim\sum_{j=0,1,2,\dots}^{\infty}\epsilon^{j}g_{j}\mathpunct{,}
\end{equation}
where $g_{0}$ is the equilibrium value of the respective variables
and $g_{j}$s are corresponding higher-order perturbations. The equilibrium
ion velocity and plasma potential are assumed to be zero and the debris
charge density is expected to contribute \emph{only} at its second
order $\rho_{2}\equiv(\rho_{{\rm deb}})_{2}$ \citep{fkdvleo1}. The
stretched coordinates used in this case are 
\begin{equation}
\zeta=\epsilon^{1/2}(x-v_{{\rm ph}}t),\quad\tau=\epsilon^{3/2}t,
\end{equation}
corresponding to the coordinate $(x,t)$, where $v_{{\rm ph}}$ is
the phase velocity of the wave. Following the standard mathematics,
an fKdV equation can be obtained \citep{fkdvleo1} in the first-order
perturbation in $\phi$ as 
\begin{equation}
\frac{\partial\phi_{1}}{\partial\tau}+\frac{1}{2}\left(\frac{\partial\phi_{1}^{2}}{\partial\zeta}+\frac{\partial^{3}\phi_{1}}{\partial\zeta^{3}}\right)=\frac{1}{2}\frac{\partial\rho_{2}}{\partial\zeta},
\end{equation}
with $\rho_{2}$ as a function of the stretched variable $\zeta+(1-v_{{\rm deb}})t$.
It can now be shown that for a Gaussian-shaped debris potential $\rho_{{\rm deb}}(\zeta)\sim Ae^{-(\zeta/\delta)^{2}}$
with $A<0$ (negative charge perturbation), the above equation results
in various kinds of pinned solitons \citep{fkdvleo1}, localized at
the site of the perturbation. Note that the higher the value of $\delta$,
the wider is the width of the debris charge distribution.

The above model can also be used to explain the formation of DSWs
in ion density through the nonlinear Schrödinger equation (NLSE).
The NLSE can be derived by using the same reductive perturbation theory
but with a different set of stretched coordinates 
\begin{equation}
\zeta=\epsilon(x-v_{{\rm ph}}t),\quad\tau=\epsilon^{2}t,
\end{equation}
and expansion of the variables $g=(n_{i},v_{i},\phi,\rho_{{\rm deb}}\mathclose{)}$
as 
\begin{equation}
g(x,t;\zeta,\tau)=g_{0}+\sum_{j=1}^{J}\epsilon^{j}\sum_{l=-L}^{L}g_{j,l}(\zeta,\tau)\,e^{il(kx-\omega t)}\mathpunct{,}
\end{equation}
where $v_{{\rm ph}}=\omega/k$ is the phase velocity as before and
the debris is supposed to contribute \emph{only} in the second order.
With these expansions, Eqs.(\ref{eq:cont-1}-\ref{eq:pois-1}) can
be reduced to an NLSE or a Gross-Piteavskii equation (GPE) in $\phi_{1,1}$
\begin{equation}
i\frac{\partial\phi_{1,1}}{\partial\tau}+C_{1}\frac{\partial^{2}\phi_{1,1}}{\partial\zeta^{2}}+C_{2}|\phi_{1,1}|^{2}\phi_{1,1}=C_{3}\rho_{2}\phi_{1,1},\label{eq:nlse}
\end{equation}
with $C_{1,2,3}$ as certain constants \citep{sarkar}. With the proper
choice of $\rho_{2}$, the above equation can be solved in $\phi_{1,1}$
or in ion density to show that it admits DSWs \citep{sarkar}.

\section{Counter-streaming ions and pinned solitons}

We note that the IICSI is the dominant one when the external charge
perturbation is negative \citep{debye-scale}. The saturation of IICSI
is caused by ion-trapping and ion-heating, both of which depend on
each other. Our simulations yield growth rates consistent with the
linear theory and phase space structures. The electron-ion streaming
instability is not relevant here as it is slower by one or more orders
of magnitude than the IICSI. This is also revealed in earlier kinetic
(Vlasov) simulation \citep{iicsi}.

We now focus on the situation of what happens, when a Debye-scale
structure (debris) of negative charge causes ions to counter-stream
and ram into one another, causing an ion-ion two-stream electrostatic
instability to develop and become turbulent when the debris charge
is considerably large. We shall show that this turbulence caused by
counter-streaming ion instability results in the formation of pinned
solitons. In what follows, we shall build up a kinetic model for this
to happen and follow it up with a PIC simulation.

\subsection{Linear theory}

Here, we are going to briefly describe the linear analysis of electrostatic
instability in a plasma with equilibrium velocities in the framework
of kinetic theory. Let us consider a multi-species, quasi-neutral,
collision-less plasma with different species, having different equilibrium
drift velocities. The basic governing equations are the electrostatic
Boltzmann-Vlasov equations for different species and Poisson's equation
for closure 
\begin{eqnarray}
\frac{\partial f_{j}}{\partial t}+\bm{u}_{j}\cdot\nabla f_{j}+\frac{q_{j}}{m_{j}}\bm{E}\cdot\frac{\partial f_{j}}{\partial\bm{u}_{j}} & = & 0,\\
\epsilon_{0}\nabla\cdot\bm{E} & = & \sum_{j}q_{j}n_{j},
\end{eqnarray}
where $(f,\bm{u},q,n)_{j}$ are the velocity distribution functions
(VDFs), velocity, charge, and number density of the $j$th species
respectively, and $\bm{E}$ is the electric field. The VDF in general,
can be expressed as a function of velocities and temperature for each
species. For Maxwellian case, it becomes 
\begin{equation}
f_{j}(u)\sim\exp\left[-\frac{(u-v_{j})^{2}}{2c_{j}^{2}}\right],
\end{equation}
where $v_{j}$ and $c_{j}$ are the equilibrium drift velocity and
thermal speed of the $j$th species, respectively, 
\begin{equation}
c_{j}=(T_{j}/m_{j})^{1/2},
\end{equation}
where the temperature is expressed in energy unit with $T_{j}$ and
$m_{j}$ being the temperature and particle mass of the $j$th species.
The quantity $c_{j}$, referred here as the thermal speed of species
`$j$' should not be confused with the ion-sound speed, which is denoted
by $c_{s}$.

We now introduce a small electrostatic perturbation and express various
physical quantities $F$ with a linear perturbation scheme, $F=F_{0}+F_{1}$,
where $F_{0,1}$ are equilibrium and perturbed parts. In our case
\begin{equation}
F=(f,\bm{E}),\quad F_{1}\sim e^{-i\omega t+i\bm{k}\cdot\bm{r}}.
\end{equation}
Further linear analysis of the above model leading to the linear dispersion
relation is quite well-known \citep{chen-1974} and the final dispersion
relation can be written as 
\begin{equation}
k^{2}-\sum_{j}\omega_{pj}^{2}\left(\mathbb{P}\int_{-\infty}^{+\infty}\frac{\hat{f}_{j0}'}{u_{j}-\omega/k}\,du_{j}+i\pi\left.\hat{f}_{j0}'\right|_{u_{j}=\omega/k}\right)=0.\label{eq:dis1-1-1}
\end{equation}
For clarity, a detailed analysis is given in Appendix A. Note that
$\hat{f}_{j0}$ is the unit-normalized velocity distribution function
{[}see Eq.(\ref{eq:unit2}){]} and 
\begin{equation}
\hat{f}_{j0}'=\frac{\partial\hat{f}_{j0}}{\partial u_{j}}.
\end{equation}

\textcolor{black}{The situation, we are particularly looking at is:
when a negatively charged external perturbation (debris) is present
in an $e$-$i$ plasma, then from both sides of the debris there will
be two counter-streaming ion populations moving towards the debris
site and two counter-streaming electron populations moving away from
it.} However, in order to simplify the mathematics, we shall consider
two drifting populations of ions with velocity $\pm v/2$. Besides,
in order to eliminate kinetic effects of the electrons, we consider
electrons to be Boltzmannian. Thus the ion VDFs are given by 
\begin{equation}
\hat{f}_{i0}(u)=\frac{1}{\sqrt{2\pi c_{i}^{2}}}\left\{ \exp\left[-\frac{(u+v/2)^{2}}{2c_{i}^{2}}\right]+\exp\left[-\frac{(u-v/2)^{2}}{2c_{i}^{2}}\right]\right\} .
\end{equation}
The above equation denotes two similar counter-streaming ion populations.
As the debris charge is a symmetric Gaussian, the counter streams
of ions are symmetric. The linearized Poisson's equation, Eq.(\ref{eq:pois})
becomes 
\begin{equation}
ik\epsilon_{0}E_{1x}=-en_{e1}+en_{i1},
\end{equation}
with 
\begin{eqnarray}
n_{e1} & = & n_{0}\left(\frac{e\phi_{1}}{T_{e}}\right),\label{eq:pert}\\
n_{i1} & = & \int f_{i1}\,d\bm{u}_{i},
\end{eqnarray}
where $f_{i1}$ is given by Eq.(\ref{eq:f1}). Eq.(\ref{eq:pert})
is the linearized perturbed electron density, while the expression
for electron density, in general, is given by 
\begin{equation}
n_{e}=n_{0}\,\exp\left(\frac{e\phi}{T_{e}}\right),
\end{equation}
Substituting these expressions in Eq.(\ref{eq:dis1-1-1}), we have
\begin{equation}
k^{2}=-\frac{1}{\lambda_{De}^{2}}+\frac{1}{2}\omega_{pi}^{2}\left(\int_{-\infty}^{+\infty}\frac{F'\left(\tilde{u}^{+}\right)}{u-\omega/k}\,du+\int_{-\infty}^{+\infty}\frac{F'\left(\tilde{u}^{-}\right)}{u-\omega/k}\,du\right),\label{eq:dis2}
\end{equation}
where we have substituted $\phi_{1}=iE_{1x}/k$ and 
\begin{equation}
F(u)=\frac{1}{\sqrt{2\pi c_{i}^{2}}}\exp\left(-\frac{u^{2}}{2c_{i}^{2}}\right),
\end{equation}
with $\tilde{u}^{\pm}=u\pm v/2$. Note that each ion population is
responsible for half of the total ion density. Expressing $u$ in
terms of $\tilde{u}$, the integrations can be written as 
\begin{equation}
\int_{-\infty}^{+\infty}\frac{F'(\tilde{u})}{\tilde{u}-\tilde{\omega}^{\pm}/k}\,d\tilde{u},
\end{equation}
where $\tilde{\omega}^{\pm}=\omega\pm kv/2$ is the Doppler-shifted
frequency. For small $\textrm{Im}(u,\tilde{u})$, one can approximate
Eq.(\ref{eq:dis2}) as 
\begin{eqnarray}
k^{2} & = & -\frac{1}{\lambda_{De}^{2}}+\frac{1}{2}\omega_{pi}^{2}\left(\mathbb{P}\int_{-\infty}^{+\infty}\frac{F'(u)}{u-\tilde{\omega}^{+}/k}\,du+\mathbb{P}\int_{-\infty}^{+\infty}\frac{F'(u)}{u-\tilde{\omega}^{-}/k}\,du\right)\nonumber \\
 &  & +\,i\pi\frac{1}{2}\omega_{pi}^{2}\left(\left.\frac{\partial F}{\partial u}\right|_{u=\tilde{\omega}^{+}/k}+\left.\frac{\partial F}{\partial u}\right|_{u=\tilde{\omega}^{-}/k}\right).\label{eq:prin}
\end{eqnarray}
Proceeding as before, we can estimate the growth rate of the instability
as 
\begin{eqnarray}
\textrm{Im}\left(\frac{\omega}{\omega_{pi}}\right) & \simeq & \frac{1}{4}\sqrt{\alpha^{3}\pi}\left(\frac{\omega_{pi}}{kc_{i}}\right)^{3}\tilde{\omega}_{pi}\left\{ \left(\frac{kv}{2\omega_{pi}}-1\right)\exp\left[-\frac{(\omega_{pi}-kv/2)^{2}}{2k^{2}c_{i}^{2}}\right]\right.\nonumber \\
 &  & \left.-\,\left(\frac{kv}{2\omega_{pi}}+1\right)\exp\left[-\frac{(\omega_{pi}+kv/2)^{2}}{2k^{2}c_{i}^{2}}\right]\right\} ,\label{eq:gam}
\end{eqnarray}
where $\alpha$ is a dimensionless parameter defined in Eq.(\ref{eq:alpha}).
The detailed analysis leading to the derivation of the above relation
is given in Appendix B.

\begin{figure}[t]
\begin{centering}
\includegraphics[width=0.5\textwidth]{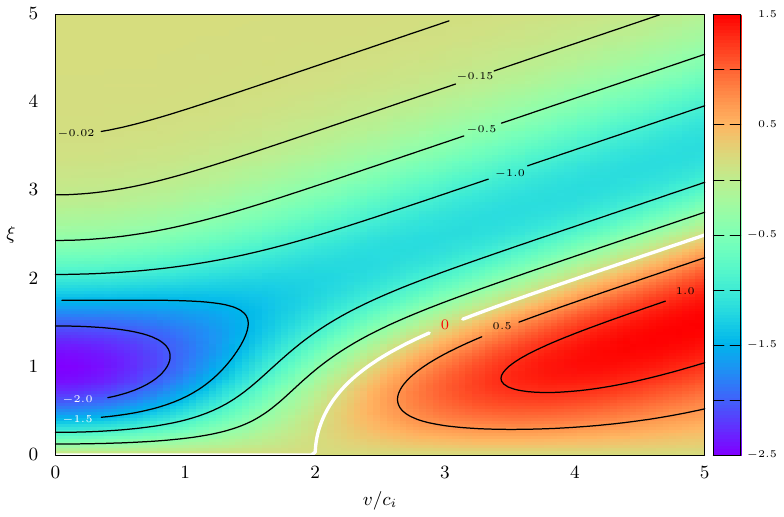} 
\par\end{centering}
\caption{\protect\protect\label{fig:Contour-plot-of}Contour plot of the
function ${\cal F}$.}
\end{figure}

\subsection{Critical velocity}

Normalizing $v\to v/c_{i}$ and $\gamma=\textrm{Im}(\omega/\omega_{pi})$,
one can conveniently express Eq.(\ref{eq:gam}) as 
\begin{equation}
\gamma=\frac{1}{8}\tilde{\omega}_{pi}\xi^{2}\sqrt{\pi\alpha^{3}}{\cal F},
\end{equation}
where $\tilde{\omega}_{pi}$ is a dimensionless ion plasma frequency
as defined in Eq.(\ref{eq:omega}) and 
\begin{equation}
{\cal F}=\left(v-2\xi\right)\exp\left[-\frac{(v/2-\xi)^{2}}{2}\right]-(v+2\xi)\exp\left[-\frac{(v/2+\xi)^{2}}{2}\right]\label{eq:f}
\end{equation}
with $\xi=1/(k\lambda_{Di})$ and $\gamma$ as the normalized growth
rate, the sign of which is determined by ${\cal F}$. A contour plot
of the term ${\cal F}$ is shown in Fig.\ref{fig:Contour-plot-of}.
As shown in the figure, one can find out the minimum critical velocity
$v_{c}$ of the ion streams required to excite the instability $(\gamma>0)$
can be calculated from the curve ${\cal F}=0$ by seeking the minimum
$v$. However, solutions of the equation ${\cal F}=0$ can only be
expressed in terms of inverse functions and in general one cannot
obtain a full spectrum of solutions including the equation for the
curve of minimum $v$ (the white curve in Fig.\ref{fig:Contour-plot-of}).
One can, however, obtain $v_{c}$ by expanding ${\cal F}$ around
$\xi=0$ and find the solution of the resultant equation after setting
$dv/d\xi=0$ as, 
\begin{equation}
\left(v_{c}^{2}-4\right)e^{-v_{c}^{2}/8}=0,
\end{equation}
which yields $v_{c}=2$. In terms of ion-sound speed $c_{s}=\sqrt{T_{e}/m_{i}}$,
$v_{c}\simeq2\sqrt{\sigma}c_{s}\sim0.63c_{s}$ for $\sigma=0.1$,
which is quite within the limit $c_{s}>v_{{\rm c}}>1.3c_{i}$ \citep{history3}.
For a situation when external charge perturbation (i.e.\ debris)
is stationary relative to the plasma, it is the debris potential which
causes acceleration of the ions. The maximum velocity $v_{\textrm{max}}$
an ion can obtain for a potential difference of $\varphi$ between
the debris and bulk plasma can be found by equating the electrostatic
energy to the kinetic energy of the ions $v_{\textrm{max}}=(2ef\varphi/m_{i})^{1/2}$,
where $f$ is the fraction of electrostatic energy that is converted
to the kinetic energy of the ions, which lies between $0$ and $1$
and $e$ is the electronic charge. Normalizing the potential by $(T_{e}/e)$
and velocity by $c_{s}$, we find $\phi=v_{\textrm{max}}^{2}/(2f)\sim2\sigma f^{-1}$.
Usually, when we consider the acceleration of charged particles through
an electrostatic potential drop, we consider the source of the potential
to be infinitely large so that all particles get accelerated equally.
However, in this case, when an ion is accelerated by the debris potential,
it will reduce the potential for the subsequent particles and the
effective potential available for incoming particles will be continuously
reduced, which is being taken care of by the factor $f$, much like
Debye shielding.

At this point, we should also note that the magnitude of $v_{c}$
is directly related to the $\rho_{{\rm deb}}$ in the sense that the
higher is the value of $\rho_{{\rm deb}}$, the larger is the value
of $v_{c}$. So, a higher $\rho_{{\rm deb}}$ should quickly trigger
the instability while for a smaller value of $\rho_{{\rm deb}}$,
the perturbation causing the instability should be completely thermalized.
We can see this effect when we discuss the onset of turbulence in
Section V, where we can see that there exists a critical value of
$\rho_{{\rm deb}}$ beyond which only turbulence can set in.

\section{PIC simulation}

We now perform a PIC simulation of a 1-D $e$-$i$ plasma with positively
and negatively charged debris. We shall see the formation of DSWs
for a strong positively charged external perturbation (debris) and
how the turbulence caused by counter-streaming ions due to a negatively
charged external perturbation (debris) gives rise to pinned solitons.
The PIC code used in these simulations is the \emph{h}-PIC-MCC code,
which can simulate plasma processes in the IA, dust-ion-acoustic (DIA),
and electron time scales with dust-charge fluctuation. More about
this code and its benchmarking results can be found in the papers
by Changmai and Bora \citep{suniti1,suniti2} and Das et al. \citep{suniti3}.
\begin{figure}[t]
\begin{centering}
\includegraphics[width=1\textwidth]{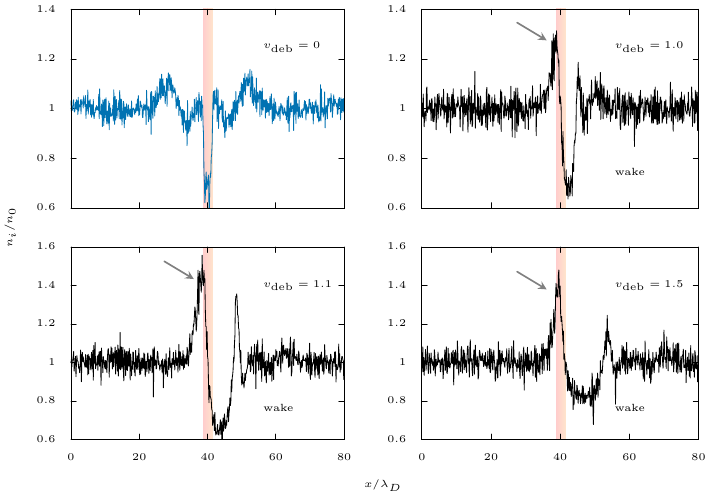} 
\par\end{centering}
\caption{\protect\protect\label{fig:Formation-of-DSW}Formation of DSW in
ion density due to a positive external charge perturbation. The vertical
colored stripe, in each panel, indicates the position of the Gaussian
shaped positive charge debris. The arrows indicate the DSW structure
as the debris moves from right to left with velocity $v_{{\rm deb}}$
(normalized by ion-sound speed $c_{s}$). For a static perturbation,
what we see is the propagation of an IAW from the site of initial
perturbation, which changes to a DSW as velocity increases. The figures
are in the rest-frame of the debris.}
\end{figure}

In our simulation, the simulation box length is $0.006\,{\rm m}$,
where both electrons and ions are represented each with $10^{5}$
computational particles with an ion-to-electron mass ratio of $1836$.
The simulation box is divided into grid of $600$ uniform cells. For
the plasma parameters used in this simulation (please see the next
subsection), the electron Debye length comes out to be $\lambda_{D}\sim7.4\times10^{-5}\,{\rm m}$
and the electron plasma frequency $\omega_{{\rm pe}}\sim5.9\times10^{9}\,{\rm rad/s}$.
We use a numerical evolution time step $\sim10^{-11}\,{\rm s}$. With
these simulation parameters, we are able to have full temporal resolution
at electron timescale and spatial resolutions over the entire simulation
domain. In our simulation, we have distributed the electrons and ions
with their respective Maxwellian velocity distributions \citep{bridsall}.
This results in a linear Poisson's equation Eq.(\ref{eq:pois-1}),
which is solved with successive over-relaxation (SOR) method. The
simulation is being run with periodic boundary conditions.

\subsection{Formation of DSW}

The results of this PIC simulation of this model of an $e$-$i$ plasma
are shown in Fig.\ref{fig:Formation-of-DSW} (this study), which shows
the formation of DSW when $v_{{\rm deb}}\geq c_{s}$ for $\rho_{{\rm deb}}>0$,
where $v_{{\rm deb}}$ is the debris velocity w.r.t background plasma
and $c_{s}=\sqrt{T_{e}/m_{i}}$ is the ion-sound speed with temperature
expressed in energy units. The debris is a Gaussian-shaped charge
distribution with width $\sim\lambda_{D}$, which is of the order
of a few Debye lengths, shown as an orange-colored strip in Fig.\ref{fig:Formation-of-DSW}.
The relevant plasma parameters in this case are as follows: plasma
number density $n_{i}\sim n_{e}=10^{16}\,{\rm m}^{-3}$, electron
temperature $T_{e}=1\,{\rm eV}$, ion temperature $T_{i}=0.01\,{\rm eV}$,
and $|\rho_{{\rm deb}}|\sim0.09\rho_{0}$ without taking into account
the background plasma density and $\rho_{0}=en_{i,e}$ is the equilibrium
plasma density and $e$ being the value of electronic charge. All
figures are drawn in the rest-frame of the debris. The plasma parameters
used in this simulation work can be representative of space plasma
such as found in low-Earth orbit (LEO) plasma, though we have used
a relatively higher plasma density so as to enhance the effect of
debris (external charge perturbation) according to $|\rho_{{\rm deb}}|\sim0.09\rho_{0}$
for more clear observation of the debris effect. Note that the instability
growth rate $\gamma\propto\tilde{\omega}_{pi}$, the re-defined ion-plasma
frequency which essentially implies that $\gamma\propto n_{i}^{1/2}$.
So the lower is the plasma density, the lower is the growth rate and
instability will take a longer time to develop. By using a bit higher
plasma density we can make the effect of the instability visible quickly
without compromising other physical effects. As the primary objective
of this work is to understand the physics of different nonlinear structures
formed due to the presence of external charged debris, we use the
results of work as a `proof of principle'. So, the results presented
are general in nature and are expected to enrich the general understanding
of such situations rather than exactly explaining specific details
of any particular situation. For example, similar phenomena are experimentally
observed in laboratory plasmas, the parameters for which are representative
of the earth's bow shock region \cite{pinned}.

As can be seen from the figure, for a static debris, the perturbation
results in the formation of IAW, which propagates away from the site
of the debris, while a dispersive (oscillating) shock front appears
on the front of the debris with an IA wake when there is a relative
velocity between the plasma and the debris. These results are in good
agreement with fluid simulation results, as reported by Sarkar and
Bora \citep{sarkar}, which also provides a purely theoretical explanation
of these oscillations in terms of nonlinear Schrödinger equation (NLSE)
in the plasma potential $\phi$ \citep{sarkar}.

\subsection{Formation of pinned solitons}

In this section, we present the results of our PIC simulation and
show how a localized counter-streaming ion instability can form, sustain
and gives rise to the formation of pinned solitons through electrostatic
turbulence. We shall also see how the simulation results closely agree
with our theoretical analysis (for the linear regime), as already
described in Section III. 
\begin{figure}
\begin{centering}
\includegraphics[width=1\textwidth]{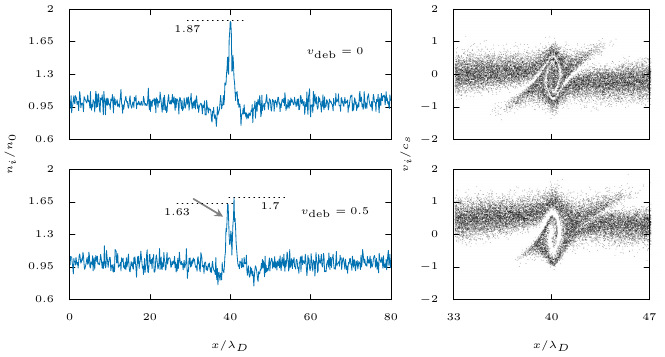}\\
 \includegraphics[width=1\textwidth]{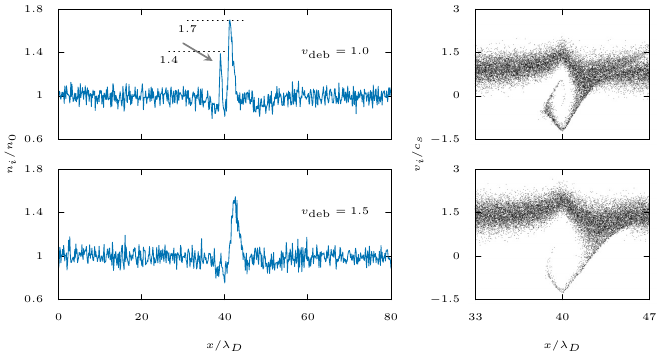} 
\par\end{centering}
\caption{\protect\protect\label{fig:Formation-of-pinned}Formation of pinned
solitons due to a negative external charge perturbation at the debris
site for different debris velocity $v_{{\rm deb}}$ (normalized by
ion-sound speed $c_{s}$). In the figure, we have shown the ion density
$n_{i}$ (left panel) and the corresponding phase-space for the pinned
solitons, showing phase-space vortices, indicating electrostatic turbulence
(right panel). The arrow shows the pinned solitons. Pinned solitons
cease to form at higher $v_{{\rm deb}}$ (lower panel).}
\end{figure}

The results of this simulation are shown in Fig.\ref{fig:Formation-of-pinned},
in which we plot the ion density $n_{i}$ as well as the scatter plot
of the ion phase-space for various $v_{{\rm deb}}$ for a negative
charge perturbation $(\rho_{{\rm deb}}<0)$ in the rest frame of the
debris, which is centered in the simulation box at $\sim40\lambda_{D}$.
What we observe is a \emph{localized} acceleration of ions by the
negatively charged debris potential causing a counter-streaming ion
instability to develop which then goes on to the nonlinear saturation
regime, ultimately entering a turbulent regime. The left panels in
the figure show the ion density, which are the pinned solitons, resulted
by trapping of ions in the phase space vortices. We further see that
as debris velocity increases, it causes the widening of the vortices
causing well-separated pinned solitons. In the limit of $v_{{\rm deb}}\to0$,
the single soliton that we can see is nothing but the merger of multiple
pinned solitons. This is evident from the fact that as debris velocity
increases, the vortices get widened and the solitons are un-merged
which ultimately gets separated. One can also see that this un-merging
causes the heights of the final solitons to decrease due to redistribution
of the energy, originally contained in a single soliton only (see
Fig.\ref{fig:Formation-of-pinned}, the heights of the solitons are
marked for easy reference). It should, however, be noted that this
is distinctly different from the situation where two independent solitons
approach and cross each other, making the composite soliton height
increase momentarily during the cross over. In our case, the multiple
solitons un-merge into separate solitons. However, when $v_{{\rm deb}}\gg c_{s}$,
the vortices are widened up to a limit where pinned solitons cannot
form. This is primarily due to the fact that `free' (linear) perturbations
will not move faster than $c_{s}$. So, pinned solitons can form \emph{only}
within a window of range of debris velocities. These results are consistent
with the findings of Tiwari \citep{fore-wake} and Sarkar \citep{sarkar}.
At this point, we would also like to emphasize that DSWs can possibly
form with positively charged debris or in other words the IICSI \emph{cannot}
be responsible for the formation of any shock structure, which also
agrees with an earlier study \citep{turb}.

\begin{figure}[t]
\begin{centering}
\includegraphics[width=0.5\textwidth]{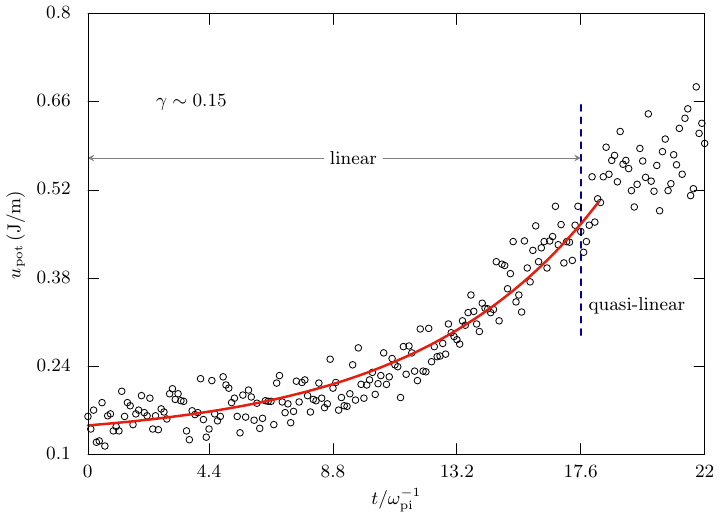}\hfill{}\includegraphics[width=0.5\textwidth]{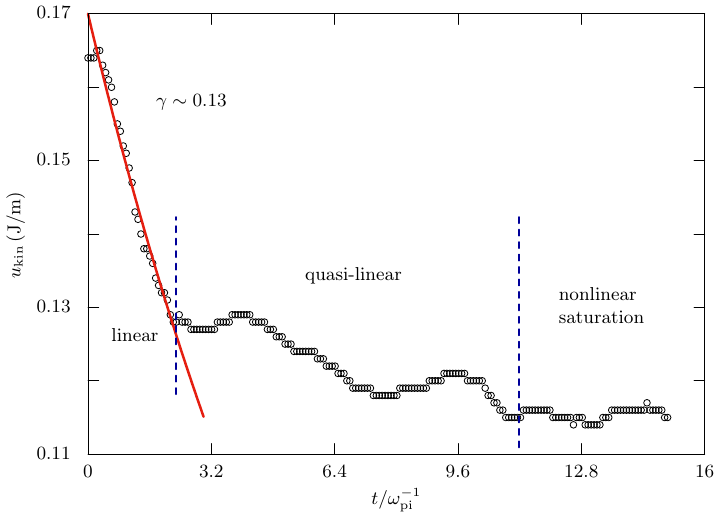} 
\par\end{centering}
\caption{\protect\protect\label{fig:Potential-(left)-and}Potential (left)
and kinetic (right) energy histories of the classical counter-streaming
ion (left) and debris-driven ion instabilities (right). Both plots
show energy densities as the instabilities develop. While the open
circles show the respective simulation data, the solid lines show
the exponential rise (decay) of the potential (kinetic) energies during
the linear regime of the instability. The other regions are marked
accordingly.}
\end{figure}

It will be interesting to compare these extremely localized counter-streaming
instabilities to the classical streaming instability. In the usual
case of streaming instability, the entire ion population rams into
one another and vortices form all over the domain, whereas in a charged
debris-induced case, as we have seen, the instability is severely
localized, only in the neighborhood of the debris, which causes the
pinned solitons to form. In Fig.\ref{fig:Potential-(left)-and} (left
panel), we show the formation of such an instability and the electrostatic
energy history of the plasma (in terms of potential energy density
$u_{{\rm pot}}$), which shows the linear growth of the instability.

The linear growth is marked by the exponential rise of the electrostatic
energy, followed by a quasilinear stage and nonlinear saturation.
The linear growth rate in this case is $\gamma\sim0.15$. In this
case, the initial kinetic energy of the counter-streaming ions is
converted to electrostatic potential energy as the instability grows.
The streaming velocity of the ions with respect to their counter-streaming
part is $\sim3c_{i}$.

On the other hand, in the debris-induced streaming instability, the
electrostatic potential energy of the debris is converted to the kinetic
energy of the accelerating ions, which is then converted back into
the electrostatic potential energy of the ions as they become trapped
in the phase space vortices. This evolution of kinetic energy is shown
in the right panel of Fig.\ref{fig:Potential-(left)-and}. Note that
the decrease of kinetic energy is at the cost of increasing electrostatic
energy which is depicted by the decay curve of the kinetic energy.
In this case, we have chosen to plot the kinetic energy as the instability
is extremely localized. We note that in PIC simulation, the electrostatic
potential is interpolated at the grid points, which makes it usually
difficult to find the localized contributions of the instability to
the increase in potential energy. However, as the kinetic energy is
calculated at the particle positions, it is rather easier to calculate
the kinetic energy for particles confined within a region and is a
good marker of the underlying instability. The situation however is
different in the case of a non-localized streaming instability, where
the potential and kinetic energies of the entire population of the
plasma particles get evolved, such as one shown in the left panel
of Fig.\ref{fig:Potential-(left)-and}. In this case, one can use
either potential or kinetic energy to keep track of the instability.
Also note that the physical and computational models do not have any
dissipative effects, making the system fully conservative and one
can use either potential or kinetic energy, depending on the ease
of use. The instability growth rate in the debris-driven case is found
to be $\gamma\sim0.13$, corresponding to the topmost plot in Fig.\ref{fig:Formation-of-pinned}
where the debris velocity w.r.t the background plasma is zero. The
effective counter-streaming velocity of the ions towards the debris
site in this case is $\sim5.6c_{i}$, which is almost twice the value
needed for classical IICSI for the equivalent growth rate. This can
be understood from the extremely localized nature of the instability.
As the plasma away from the site of perturbation (debris) is \emph{not}
affected, the surrounding plasma tends to thermalize the perturbation
requiring a higher streaming velocity to sustain the instability.
It is also obvious why these events are extremely localized, which
is due to the Debye shielding of the disturbance caused by debris.

Before we proceed further, we would like to discuss the exact nature
of external charged debris in a flowing plasma. In situations such
as LEO, charged debris may arise due to the presence of different
kinds of so-called foreign objects such as parts of de-commissioned
and broken parts of satellites, collectively referred to as `space
debris' \cite{perek}. These debris may get charged due to different
reasons \cite{garrett,whipple,havness,goree,juhasez,anderson} including
photoemission of electrons as a result of exposure to high energy
solar radiation \cite{mandell}. We should also note that the cumulative
amount of charge of a debris is not expected to remain constant in
time but should vary with different timescales and may also have a
stochastic component of variability as well \cite{vaulina}. However,
modeling such variations is extremely difficult as we do not have
enough first-hand experimental data on such variations. This is why,
in all such work dealing with external charged debris in flowing plasmas,
it is customary to treat the debris charge as constant in time. In
this PIC simulation also, we have treated the cumulative charge and
shape of the debris constant in time. Dynamically, a PIC plasma simulation
is quite closer to a realistic physical process, so long as the plasma
is not strongly coupled. So, when an external debris is introduced
into a plasma, the surrounding plasma will try to neutralize the excess
charge, much like a Debye shielding effect, as already mentioned in
Section III. In the absence of any relative velocity between the debris
and the bulk plasma, the localized disturbance created by the charged
debris is expected to thermalize as time progresses. For the parameters
used in this simulation, we see that the fastest ion-acoustic time
scale $\tau_{{\rm IA}}\sim(kc_{s})^{-1}$, assuming the largest $k\sim2\pi/(10\lambda_{D})$
with $10\lambda_{d}$ as the average size of debris, is of the same
order as the ion plasma timescale $\tau_{pi}$. On the other hand,
the IICSI growth rate is quite slower $\sim0.1\omega_{pi}$ and the
thermalization time should be about several order magnitudes higher
than the IICSI timescale. As a result, when there is a relative velocity
and the debris becomes a moving object and the localized disturbance
never gets time to thermalize, as by the time the instability builds
up, the debris moves ahead and the IICSI continues to build up keeping
the structures sustained in time, which is what we have seen in the
simulation.

\section{Onset of turbulence}

We shall now try to see the onset of turbulence due to an external
charge perturbation or debris. However, an interpretation of 1D turbulence
is in order here. Heuristically, any discussion on turbulence essentially
leads to Kolmogorov's 1941 work \citep{kol1}, which provides a scaling
on energy cascades in an inertial subrange 
\begin{equation}
E(k)\sim\varepsilon^{2/3}k^{-5/3},
\end{equation}
where $E(k)$ is the turbulent energy density, $k$ is the wave number,
and $\varepsilon$ is the energy available for the fluid per unit
mass and time. Alternatively, $\varepsilon$ can also be interpreted
as the energy which cascades from scale to scale and gets dissipated
by viscosity beyond the inertial subrange. However, the above scaling
is true strictly for a three-dimensional situation whereas our observation
is for a dissipation-less 1D case with dispersion. Therefore, we identify
our case with the basic and well-known 1D turbulence model by Majda,
McLaughlin, and Tabak \citep{mmt} (thereafter MMT) which describes
the 1D turbulence for dispersive wave with the help of a family of
dynamical equations \citep{zakh} 
\begin{equation}
i\frac{\partial\psi}{\partial t}=\left|\frac{\partial}{\partial x}\right|^{\eta}\psi+\lambda\left|\frac{\partial}{\partial x}\right|^{\beta/4}\left(\left|\left|\frac{\partial}{\partial x}\right|^{\beta/4}\psi\right|^{2}\left|\frac{\partial}{\partial x}\right|^{\beta/4}\psi\right),\quad\lambda=\pm1,\label{eq:dynamic}
\end{equation}
where $\psi(x,t)$ is a complex wave field and $(\eta,\beta)$ are
real parameters. For $\eta=2$ and $\beta=0$, the above equation
reduces to an NLSE, as given by Eq.(\ref{eq:nlse}), which is also
as in our case. Eq.(\ref{eq:dynamic}) conserves two integrals, namely
$L^{2}$-norm (wave action) and momentum 
\begin{equation}
\int|\psi|^{2}\,dx,\,{\rm and}\,\int(\psi\bar{\psi}_{x}-\psi_{x}\bar{\psi})\,dx,
\end{equation}
respectively, where the subscript denotes spatial partial derivative
and the $\bar{\psi},\bar{\psi}_{x}$ denote the corresponding complex
conjugates. Kolmogorov-type energy scaling laws are basically the
equilibrium solutions of the model, namely \citep{mmt,zakh} 
\begin{eqnarray}
\tilde{n}(\omega) & \sim & \omega^{(-2\beta/3-1+\eta/3)/\eta},\label{eq:law1}\\
\tilde{n}(\omega) & \sim & \omega^{-(2\beta/3+1)/\eta},\label{eq:law2}
\end{eqnarray}
where the quantity $\tilde{n}$ can be interpreted as the spectral
density of the wave field in $\omega$-space. Apart from these one-dimensional
energy scaling, the resultant NLSE can also support solitons and quasi-solitons
(solitons with finite lifetime) \citep{zakh}. For the parameter $\lambda=-1$,
the NLSE is known as focusing NLSE, which can support solitons and
quasi-solitons \citep{zakh}. 
\begin{figure}[t]
\begin{centering}
\includegraphics[width=0.5\textwidth]{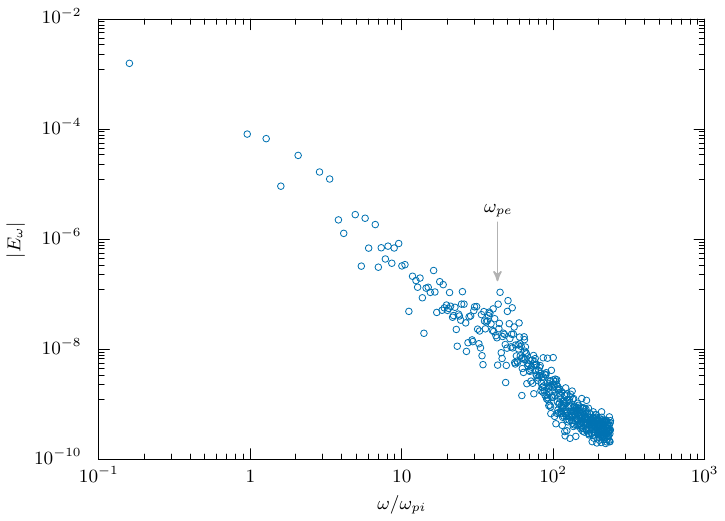}\hfill{}\includegraphics[width=0.5\textwidth]{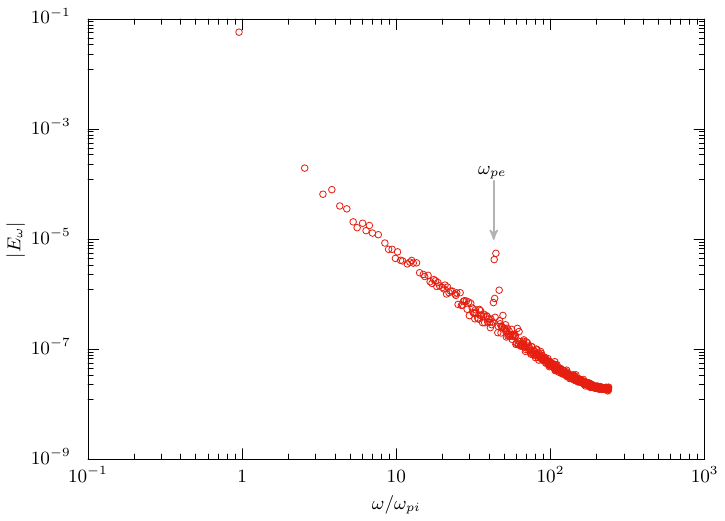} 
\par\end{centering}
\caption{\protect\protect\label{fig:The-power-spectrum}The power spectrum
density (PSD) for random fluctuations (left) and a fully developed
turbulence (right).}
\end{figure}

Following the above discussion, we now argue that the power spectrum
of the turbulent energy density in the debris-induced streaming instability
should follow similar laws like relations (\ref{eq:law1},\ref{eq:law2}).
Apparently, we can estimate the parameters $(\eta,\beta)$ from our
obtained scaling. Computationally, in order to quantify the onset
of turbulence, we plot the power spectrum density (PSD) $|E_{\omega}|$
of the kinetic energy, against frequency $\omega$. Here, $E_{\omega}$
is obtained through the Fourier transformation of the total kinetic
energy $E(t)$. The results are shown in Fig.\ref{fig:The-power-spectrum}.
In the figure, we have shown the PSDs -- one for random fluctuations
in the plasma when no particular perturbation is present (left panel)
and one when an external negative charge perturbation is introduced,
resulting in a fully developed turbulence (right panel). As is evident
from the figure, we can clearly see a Kolmogorov-type scaling in the
right panel, signifying turbulence. The relevant parameters for a
1D plasma are as follows: the equilibrium plasma density $n_{0}\sim10^{16}\,\textrm{m}^{-3}$
and electron temperature $T_{e}\sim1\,\textrm{eV}$ with $\sigma=0.01$.
Assuming that a fully developed turbulence results in a Kolmogorov-type
scaling, we try to detect the onset of turbulence by estimating the
deviation of the PSD from that of Kolmogorov-type scaling, when the
strength of the external charge debris exceeds a certain threshold.
The PSDs shown in these figures are due to the turbulence energy density
measured when the turbulence is fully developed and the instability
reaches a nonlinear saturation regime (unless otherwise stated), corresponding
to the `third region', shown in the second panel of Fig.\ref{fig:Potential-(left)-and}.

\begin{figure}[t]
\begin{centering}
\includegraphics[width=0.5\textwidth]{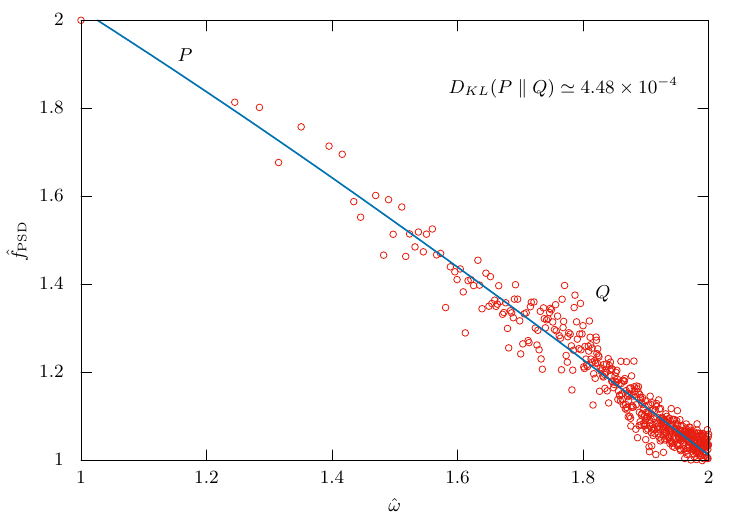}\hfill{}\includegraphics[width=0.5\textwidth]{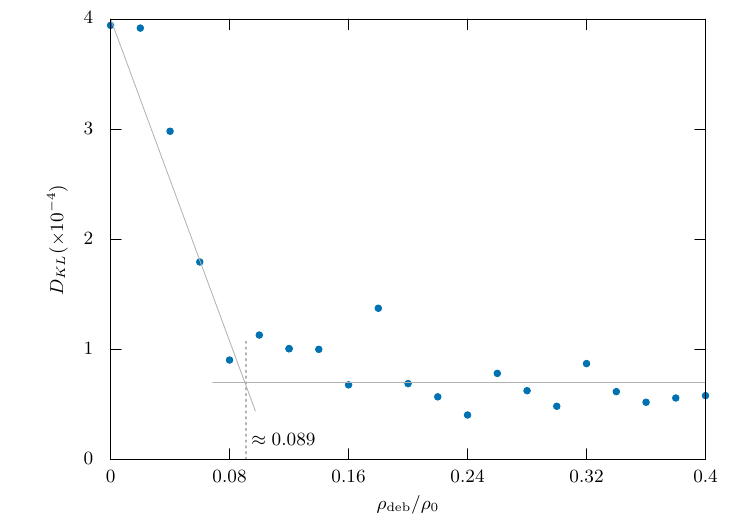} 
\par\end{centering}
\caption{\protect\protect\label{fig:The-divergence-}The divergence $D_{KL}(P\parallel Q)$
for random fluctuation is shown in the left panel. Note that both
the axes are normalized to the interval $[1,2]$ with a suitable normalized
frequency $\hat{\omega}$. A plot of $D_{KL}(P\parallel Q)$ versus
external charge density $\rho_{\textrm{deb}}$ normalized to equilibrium
charge density is shown in the right panel. The sudden change of behavior
of $D_{KL}(P\parallel Q)$ signifies the onset of turbulence. The
threshold value of $\rho_{\textrm{deb}}$ required to sustain the
instability is found to be $\sim0.089$, which is very close to the
theoretically estimated value.}
\end{figure}

In order to carry out the computational analysis, we estimate the
so-called \emph{divergence} of a dataset for PSD from the ideal one
(Kolmogorov-type) by calculating the relative entropy, also known
as Kullback--Leibler divergence, denoted as $D_{KL}(P\parallel Q)$.
This method treats the datasets as distributions $P$ (considered
as an ideal power-law distribution) and $Q$ (PSD distribution obtained
from simulation) and estimates the divergence of the target distribution
$Q$ from the sample distribution $P$. The lower is the value of
the divergence, the more closer is the distribution $Q$ to $P$.
To start with, we generate the ideal Kolmogorov-type distribution
$P$ by fitting the following nonlinear function to the distribution
$Q$, 
\begin{equation}
P\equiv\hat{f}_{\textrm{PSD}}=a+bp^{c},
\end{equation}
in the $\log$-$\log$ space. In the above fitting, $p$ denotes a
point on the PSD spectrum and $a,b,c$ are fitting constants. This
fitting is inspired by the Kolmogorov-type cascading power law. It
is to be noted that in order to avoid complex singularity in the dataset,
we have normalized the target dataset to a positive definite interval
of $[1,2]$ on both axes, although in principle one can normalize
it to any arbitrary interval. We note that the normalization is usually
desirable to avoid simultaneous appearance of both large and small
numbers, as with any other computational methods. The results of the
analysis are summarized through Fig.\ref{fig:The-divergence-}, where
we have shown the calculation of this divergence for random fluctuation
(left), for which $D_{KL}(P\parallel Q)\simeq4.48\times10^{-4}$.
On the right panel, we have plotted the divergence $D_{KL}(P\parallel Q)$
for various negative debris charge density $\rho_{\textrm{deb}}$
(external perturbation). From the figure, what we see is a sudden
shift in the value of the divergence from a linear decreasing trend
to a near-constant state, signifying the onset of turbulence. The
point of onset can be found from the intersection of the two curves
as shown in the figure, which comes out to be $|\rho_{\textrm{deb}}|\sim0.09$,
normalized to the equilibrium charge density which is unity. The random
fluctuations represent the energy cascade when no specific perturbation
is present. Since, a minimum critical velocity $v_{c}$ is required
for the counter-streaming ions for the IICSI to develop, the magnitude
of $v_{c}$ is however dependent on $\rho_{deb}$ in the sense that
a minimum $\rho_{deb}$ is required to make $v_{c}$ exceed that critical
value. Once $\rho_{deb}$ is beyond this threshold, then the higher
is the magnitude of $\rho_{deb}$ , the quicker $v_{c}$ attains its
value, and the quicker the IICSI develops.

The minimum critical velocity of the counter-streaming ions required
to sustain a turbulent regime deserves some comments before estimating
it. We should note that the critical velocity (of counter-streaming
ions) estimated through the linear analysis in Section III-B is the
one which is needed for the instability to grow. However, when the
counter-streaming instability is induced by a negative external charge
perturbation, as we have seen, the instability is highly localized.
For the instability to grow linearly and reach a nonlinear saturation,
it has to compete against the thermal fluctuations of the surrounding
plasma, before the latter can overwhelm the instability. Naturally,
it requires a higher critical velocity for the instability to be sustained.

\subsection{Estimating $v_{c}$ from the simulation}

We now estimate the critical velocity required by counter-streaming
ions to develop and sustain the instability as indicated previously.
Consider the normalized 1D Poisson's equation for debris 
\begin{equation}
\frac{d^{2}\phi}{dx^{2}}=\rho_{{\rm deb}}(x),\label{eq:pois-2}
\end{equation}
where $\rho_{{\rm deb}}(x)$ is the debris charge density, which is
a very steep-flat Gaussian-shaped, just like a top-hat function 
\begin{equation}
\rho_{{\rm deb}}(x)=\rho_{{\rm peak}}e^{-\varrho x^{2\nu}},\label{eq:rdeb}
\end{equation}
where $\rho_{{\rm peak}}$ is the value of the peak charge density
(unsigned) at $x=0$, $\varrho$ is a positive constant which essentially
determines the spread (width) of the charge distribution, and $\nu$
is a large positive integer which determines the \emph{steepness}.\emph{
}Note that here we have used a generalized expression for the symmetric
debris charge profile, though in the simulation we have used a pure
Gaussian profile. The potential should also be symmetric around the
middle point. In the extreme case, $\rho_{{\rm deb}}(x)$ becomes
a top-hat function which can be represented with a Heaviside step
function $H(x)$ 
\begin{equation}
\rho_{{\rm deb}}(x)=\rho_{0}[1-H(|x|-\varrho)].
\end{equation}
Eq.(\ref{eq:pois-2}) can then be solved analytically with the boundary
condition $(\phi,\phi')|_{x\to z}=0$ with $z>\delta L$, where $\delta L$
is the length over which the debris charge is spread out. The solution
for $\phi_{0}$ for the form of $\rho_{{\rm deb}}$ given by Eq.(\ref{eq:rdeb}),
can be written in terms of the exponential integral ${\rm Ei}_{\mu}(z)$
with 
\begin{equation}
\phi_{0}\equiv\phi|_{x=0}=\frac{\rho_{0}}{2\nu}z^{2}\,{\rm Ei}_{\mu}(\varrho z^{2\nu}),\quad\mu=1-\frac{1}{\nu},\label{eq:phi0}
\end{equation}
where $\phi_{0}$ is the value of the potential at the center of charge
perturbation and ${\rm Ei}_{\mu}(z)$ is defined as 
\begin{equation}
{\rm Ei_{\mu}(z)=}\int_{1}^{\infty}t^{-\mu}e^{-zt}\,dt,\quad{\rm Re}(z)>0.
\end{equation}
\begin{figure}[t]
\begin{centering}
\includegraphics[width=0.5\textwidth]{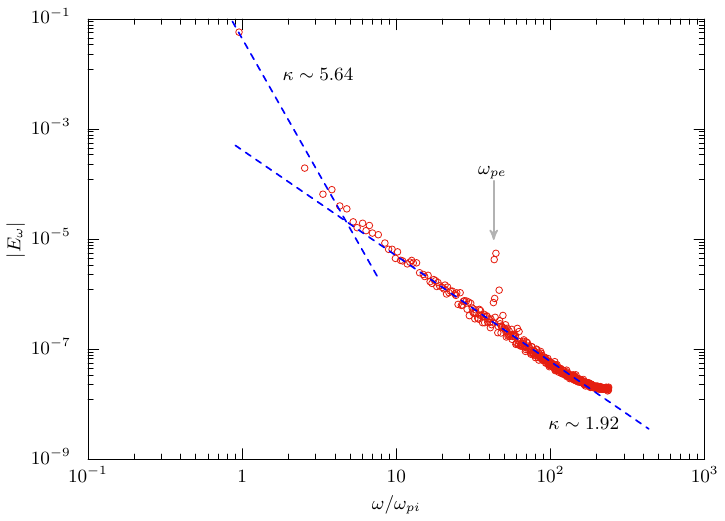} 
\par\end{centering}
\caption{\protect\protect\label{fig:Kolmogorv-type-power-laws}Kolmogorov-type
power law $(E_{\omega}\propto\omega^{-\kappa})$ for the ion-ion counter-streaming
electrostatic turbulence.}
\end{figure}

The numerical weight $w$ (which is the actual number of real plasma
particles represented by one computational particle) for the debris
in the PIC simulation is same as for electrons, as debris can be thought
to be a collection of electrons in this case. As such 
\begin{equation}
w=n_{0}L/N,
\end{equation}
where $n_{0}$ is the equilibrium plasma density $\sim10^{16}\,{\rm m}^{-3}$,
$L$ is the simulation box length which is $0.006\,{\rm m}$, and
$N$ is the number of computational particles which is $10^{5}$.
These parameters yield $w\simeq6\times10^{8}$. So, the debris charge
density $\rho_{{\rm deb}}$ can be written as 
\begin{equation}
\rho_{{\rm deb}}\sim\frac{Q_{{\rm deb}}}{\delta L},
\end{equation}
where $Q_{{\rm deb}}=ewN_{{\rm deb}}$. From our simulation, we have
found that we need about $N_{{\rm deb}}\sim800$ number of excess
computational particles to simulate the debris with $\delta L\sim(5-10)\lambda_{D}$,
required for the instability to develop and sustain. With this, we
can estimate the peak charge density $|\rho_{{\rm peak}}|\sim0.08$,
normalized to the equilibrium charge density $\rho_{0}=|en_{0}|$,
without taking account of the background plasma density. This value
can be compared with the numerically obtained one $\sim0.089$, to
which it indeed agrees very closely.

As the potential has to be symmetric around the midpoint $(x=0)$,
for $\varrho$ we can take only half of the width of the charge distribution
starting from its middle, i.e. $\delta L/2$ and can estimate from
Eq.(\ref{eq:phi0}) that $|\phi_{0}|\sim0.16$. Assuming that this
peak potential to be responsible for ion acceleration, the critical
velocity can be calculated as 
\begin{equation}
v_{c}=\sqrt{2|\phi_{0}|}\simeq0.57.
\end{equation}
Note that the above value is normalized to the ion-sound speed $c_{s}=\sqrt{T_{e}/m_{i}}$,
which when normalized by the ion thermal velocity, becomes $\sim5.65$
for $\sigma=0.01$ for which the simulation is being run. As we have
already discussed, the theoretical value calculated from the linear
theory comes out to be $2$ (see Section III-B), which is lower than
what is obtained from the simulation. This discrepancy can be justified
considering the fact that the theoretical value is correct \emph{only}
up to the order as the resultant theoretical expression Eq.(\ref{eq:f})
is heavily approximated at various levels.

It is also interesting to see the scaling law of this turbulence,
which is shown in Fig.\ref{fig:Kolmogorv-type-power-laws}. The energy-wave
number scaling $E_{\omega}\propto\omega^{-\kappa}$ ranges from a
very steep slope with $\kappa\sim5.64$ to about $\kappa\sim1.92$.
For the scaling law (\ref{eq:law2}), a probable set of parameters
{[}please see Eq.(\ref{eq:dynamic}){]} comes out to be $(\eta,\beta)\simeq(0.52,0)$.
Similar scaling is also found to be obeyed in weakly ionized plasmas
by other authors \citep{kol2}.

\subsection{Dust effects}

We end this discussion by incorporating the effect of negatively charged
dust particles on pinned solitons. This also supports our observation
of the formation of pinned solitons as results of turbulence produced
by counter-streaming ions as a response to negatively charged debris.
We should however note that studying the effect of dust particles
on the IICSI is \emph{not} among our primary objectives. We include
this section, just as a proof of concept and show that the simulation
results are consistent with the present knowledge of dust effects
on such solitons.

\begin{figure}[t]
\begin{centering}
\includegraphics[width=1\textwidth]{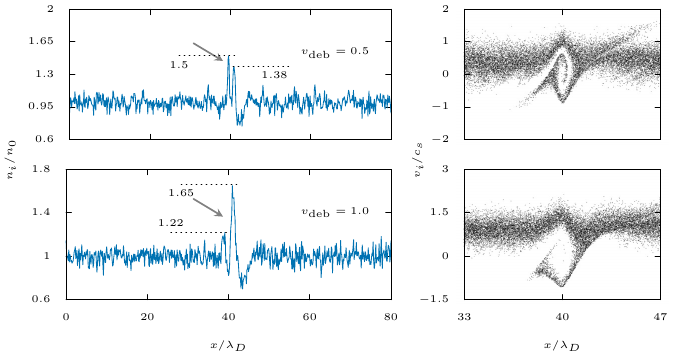} 
\par\end{centering}
\caption{\protect\protect\label{fig:Effect-of-negatively}Effect of negatively
charged dust particles on the amplitude of pinned solitons, corresponding
to the same cases for $v_{{\rm deb}}=0.5,1.0$ (normalized by ion-sound
speed $c_{s}$) without any dust (see corresponding panels in Fig.\ref{fig:Formation-of-pinned}).
The reduced amplitude of pinned solitons is clearly visible. Note
that the equilibrium dust density is kept at $0.01\%$ of the plasma
density $(n_{d0}\sim10^{-4}n_{0})$ and in the bottom panel, the increase
of height of the soliton in the right is compensated by the decrease
of height of the soliton in the left, making the total energy almost
same.}
\end{figure}

As shown in many other works \citep{sarkar}, the presence of negatively
charged dust particles increases the effective ion-sound speed of
plasma making near-sonic events sub-sonic. In the limit of large-wavelength
perturbation $(k\lambda_{D}\gg1)$ and negligible ion temperature
$(\sigma\ll1)$, the IA dispersion relation becomes \citep{sarkar}
\begin{equation}
\omega\simeq k\left(\frac{n_{i0}}{n_{e0}}\right)^{1/2}c_{s}.\label{eq:ia}
\end{equation}
Note that in the presence of negative dust particles $n_{i0}/n_{e0}>1$
due to depletion of electrons and the overall effect can be viewed
as an \emph{effective} increase of sound speed $c_{{\rm effective}}$,
with 
\begin{equation}
c_{{\rm effective}}=\left(\frac{n_{i0}}{n_{e0}}\right)^{1/2}c_{s}.
\end{equation}
This effect can be clearly seen in Fig.\ref{fig:Effect-of-negatively},
where we have included negatively charged dust particles. The dust
particles are assumed to be cold and the number density is assumed
to be constant at $10^{12}\,{\rm m}^{-3}$. The \emph{h}-PIC-MCC code
consistently takes the dust-charging as well as dust-charge fluctuation
into account \citep{suniti2,suniti3}. In Fig.\ref{fig:Effect-of-negatively},
one can directly compare the corresponding results for zero-dust cases
in Fig.\ref{fig:Formation-of-pinned} (second and third panels from
the top) where we see the reductions of amplitudes of pinned solitons
for $v_{{\rm deb}}=0.5$ and $1.0$. Note that in the bottom panel
of Fig.\ref{fig:Effect-of-negatively}, the increase of height of
the soliton in the right is compensated by the decrease of height
of the soliton in the left, making the total energy almost same. 

\section{Summary and conclusions}

In summary, we in this work, have considered an $e$-$i$ plasma in
the presence of an external, gaussian-shaped, charge perturbation
(debris) moving through the plasma in an ion-acoustic time scale.
We find that the response of the plasma differs significantly depending
on the nature and the magnitude of debris charge density and its velocity.
The simulation is being carried out with the well-tested \emph{h-}PIC-MCC
code \citep{suniti1,suniti2,suniti3}, which can take into account
dust and dust-charge fluctuation self-consistently. We have shown
that while a positively charged external perturbation produces DSWs
in the precursor region, a negatively charged perturbation causes
an IICSI, which quickly becomes turbulent, giving rise to the pinned
solitons. In the ion density plot as well as in the scatter plot of
the ion phase-space for various debris velocities, we see that when
debris velocity increases, it causes widening of the phase space vortices
causing well-separated pinned solitons, which merge to form one single
soliton when debris velocity reduces to zero. In the opposite extreme,
when debris velocity becomes highly supersonic, the vortices are widened
up to a limit causing the pinned solitons to disappear altogether.

We further show that as demonstrated through a linear analysis, the
counter-streaming ions must exceed a critical velocity $v_{c}$ in
order for the IICSI to be excited, which in this case is directly
related to the strength of negatively charged debris charge density
$\rho_{deb}$, which causes the ions to counter stream. So, in order
to have pinned soliton formation, the $\rho_{deb}$ must be above
a certain value. This value as determined from the simulation closely
agrees with the analytical estimated value. Beyond this value of $\rho_{deb}$,
the higher is the value of $\rho_{deb}$, the quicker the IICSI develops.

Through this work, we have shown that the pinned solitons are actually
manifestation of the ion phase-space vortices formed in the turbulent
regime of the ion-ion counter-streaming instability (IICSI), where
ions are effectively trapped in the potential structure. Our simulation
results are supported by the linear kinetic theory, through which
we have shown the existence of critical debris charge density for
the instability to turn turbulent. We have shown that a $E_{\omega}\propto\omega^{-\kappa}$,
Kolmogorov-type energy cascading scaling \citep{kol1,kol2} exists
in the turbulent regime which supports the formation of pinned solitons.
In this context, we have the validity of MMT model in the case of
1D turbulence which also supports the scaling law for energy cascading.
Toward the end, we have shown the effect of negatively charged dust
particles on the pinned solitons, which causes the amplitude of the
solitons to decrease and thus requires relatively large debris velocity
to make the pinned solitons appear as compared to the case when there
is no dust particle. These results largely agree with the fluid simulation
results \citep{sarkar}.

As a concluding remark, we would like to emphasize a bit about the
future outlook of this work. We have already mentioned that through
this work we have tried to look into the basic physics involving the
fundamental problem of plasma interactions with embedded charged debris.
Naturally, it is quite tempting to see whether these findings can
be extrapolated toward certain applicability namely the detection
of space debris. However, at this point, it is worth noting that the
science of detection of space debris through possible plasma activities
is still an open question and research in this direction is in a nascent
stage. Though there are other theoretical works involving plasma interaction
with external charged debris \cite{truitt}, the scientific community
as a whole, has not been able to detect such plasma activities so far,
resulting out of space debris \cite{bern} and we need more insights
into the problem including the effect of stochastic and periodic variations
of cumulative charge of such debris on plasma waves, which are some
possible interesting extensions to this work. Toward this, we believe
this work provides a renewed look into the fundamental nature of such
problems.

\section*{Acknowledgement}

One of the authors MPB would like to acknowledge DST-SERB financial
grant (CRG/2018/002971), India during which parts of this work related
to the development of the simulation code \emph{h}-PIC-MCC, was carried
out. The authors would like to thank the anonymous referees for their
constructive suggestions.

\section*{Appendix A}

\global\long\def\theequation{A\arabic{equation}}%
\setcounter{equation}{0}

The Boltzmann-Vlasov equation in the first-order perturbed distribution
function is given by 
\begin{equation}
\frac{\partial f_{j1}}{\partial t}+\bm{u}_{j}\cdot\nabla f_{j1}+\frac{q_{j}}{m_{j}}\bm{E}_{1}\cdot\frac{\partial f_{j0}}{\partial\bm{u}_{j}}=0.
\end{equation}
Without loss of any generality, we assume the perturbation to be in
$\hat{\bm{x}}$ direction so that the perturbed VDF can be written
as 
\begin{equation}
f_{j1}=-\frac{iq_{j}}{m_{j}}E_{1x}\left(\frac{\partial f_{j0}/\partial u_{j}}{\omega-ku_{j}}\right),\label{eq:f1}
\end{equation}
where $u_{j}\equiv\bm{u}_{j}\cdot\hat{\bm{x}}$. Using the linearized
Poisson's equation, we finally have 
\begin{equation}
\epsilon_{0}\nabla\cdot\bm{E}_{1}=\sum_{j}q_{j}n_{j1},\label{eq:pois}
\end{equation}
where $n_{j1}$ is the perturbed number density of the $j$th species,
expressed as a \emph{fluid} quantity, through the perturbed VDF, 
\begin{equation}
n_{j1}=\int f_{j1}\,d\bm{u}_{j},
\end{equation}
where the integration has to be carried over the 3-dimensional velocity
space.

We now write the VDF $f_{j}$ in terms of the unit-normalized VDF
$\hat{f}_{j}$, so that 
\begin{equation}
f_{j0}\equiv n_{j0}\hat{f}_{j0}.\label{eq:unit}
\end{equation}
The one-dimensional normalized VDF for $j$th species can be written
as 
\begin{equation}
\hat{f}_{j0}(u)=\left(2\pi c_{j}^{2}\right)^{-1/2}\exp\left[-\frac{(u-v_{j})^{2}}{2c_{j}^{2}}\right],\quad\int\hat{f}_{j0}\,d\bm{u}=1.\label{eq:unit2}
\end{equation}
Substituting all the expressions in the perturbed Poisson's equation,
Eq.(\ref{eq:pois}), we finally arrive at the linear dispersion equation
\begin{equation}
k^{2}-\sum_{j}\omega_{pj}^{2}\int_{-\infty}^{+\infty}\frac{\hat{f}_{j0}'}{u_{j}-\omega/k}\,du_{j}=0,\label{eq:dis1}
\end{equation}
where $\omega_{pj}$ is the plasma frequency of the $j$th species
and 
\begin{equation}
\hat{f}_{j0}'=\frac{\partial\hat{f}_{j0}}{\partial u_{j}}.
\end{equation}
We know that the integral in the above dispersion relation is singular,
which leads to the well-known Landau damping term in a plasma with
no equilibrium drift velocities. In any case, the singular integral
can be evaluated approximately assuming that the singularity lies
very close to the real $u$-axis \citep{chen-1974}. Decomposing the
integral into a non-singular part in terms of Cauchy principal value
integration and the approximating singular part through residue theorem,
we have, 
\begin{equation}
\int_{-\infty}^{+\infty}\frac{\hat{f}_{j0}'}{u_{j}-\omega/k}\,du_{j}\simeq\mathbb{P}\int_{-\infty}^{+\infty}\frac{\hat{f}_{j0}'}{u_{j}-\omega/k}\,du_{j}+i\pi\left.\hat{f}_{j0}'\right|_{u_{j}=\omega/k},
\end{equation}
where $\mathbb{P}$ indicates the principal value integration. The
final dispersion relation is now given as 
\begin{equation}
k^{2}-\sum_{j}\omega_{pj}^{2}\left(\mathbb{P}\int_{-\infty}^{+\infty}\frac{\hat{f}_{j0}'}{u_{j}-\omega/k}\,du_{j}+i\pi\left.\hat{f}_{j0}'\right|_{u_{j}=\omega/k}\right)=0.\label{eq:dis1-1}
\end{equation}

\section*{Appendix B}

\global\long\def\theequation{B\arabic{equation}}%
\setcounter{equation}{0}

The principal value integrations given in Eq.(\ref{eq:prin}) can
be found out to be 
\begin{eqnarray}
\mathbb{P}\int_{-\infty}^{+\infty}\frac{F'(u)}{u-\tilde{\omega}^{+}/k}\,du & \simeq & \frac{k^{2}}{(\omega+kv/2)^{2}}\left\langle \left(1-ku/\tilde{\omega}^{+}\right)^{-2}\right\rangle ,\\
\mathbb{P}\int_{-\infty}^{+\infty}\frac{F'(u)}{u-\tilde{\omega}^{-}/k}\,du & \simeq & \frac{k^{2}}{(\omega-kv/2)^{2}}\left\langle \left(1-ku/\tilde{\omega}^{-}\right)^{-2}\right\rangle ,
\end{eqnarray}
where $\tilde{\omega}^{\pm}/k$ are the phase velocities. The angular
brackets indicate the average of the quantity with respect to the
equilibrium VDFs $F(u)$. The quantities inside the $\left\langle \right\rangle $
provide the correction terms to the ion-plasma oscillation frequency
and to the first order, which can be neglected if we ignore its effect
on counter-streaming instability. Approximating $\left\langle \right\rangle \sim1$,
we have 
\begin{equation}
\left(1+\frac{1}{k^{2}\lambda_{De}^{2}}\right)-i\pi\frac{1}{2k^{2}}\omega_{pi}^{2}\left(\left.\frac{\partial F}{\partial u}\right|_{u=\tilde{\omega}^{+}/k}+\left.\frac{\partial F}{\partial u}\right|_{u=\tilde{\omega}^{-}/k}\right)\simeq2\frac{\omega_{pi}^{2}}{\omega^{2}}\left[\frac{1}{(2+kv/\omega)^{2}}+\frac{1}{(2-kv/\omega)^{2}}\right].
\end{equation}
Multiplying the whole equation by $\omega^{2}$ and thereafter approximating
$\omega\sim\omega_{pi}$ on the right hand side, we can simplify the
above equations as 
\begin{equation}
\omega^{2}\simeq2\alpha\omega_{pi}^{2}\tilde{\omega}_{pi}^{2}\left[1-i\pi\frac{1}{2k^{2}}\alpha\omega_{pi}^{2}\left(\left.\frac{\partial F}{\partial u}\right|_{u=\tilde{\omega}^{+}/k}+\left.\frac{\partial F}{\partial u}\right|_{u=\tilde{\omega}^{-}/k}\right)\right]^{-1}
\end{equation}
where 
\begin{eqnarray}
\alpha & = & \left(1+\frac{1}{k^{2}\lambda_{De}^{2}}\right)^{-1},\label{eq:alpha}\\
\tilde{\omega}_{pi}^{2} & = & \frac{1}{(2+kv/\omega_{pi})^{2}}+\frac{1}{(2-kv/\omega_{pi})^{2}}\label{eq:omega}
\end{eqnarray}
For small $\textrm{Im}(\omega)$, one can expand the above expression
to get an expression for the growth rate as 
\begin{equation}
\textrm{Im}(\omega)\simeq\pi\left(\frac{\alpha}{2}\right)^{3/2}\frac{\omega_{pi}^{3}}{k^{2}}\tilde{\omega}_{pi}\left(\left.\frac{\partial F}{\partial u}\right|_{u=\tilde{\omega}^{+}/k}+\left.\frac{\partial F}{\partial u}\right|_{u=\tilde{\omega}^{-}/k}\right).
\end{equation}
We note that 
\begin{eqnarray}
\left.\frac{\partial F}{\partial u}\right|_{u=\tilde{\omega}^{+}/k} & = & -\frac{(\omega/k+v/2)}{c_{i}^{3}\sqrt{2\pi}}\exp\left[-\frac{(\omega/k+v/2)^{2}}{2c_{i}^{2}}\right],\\
\left.\frac{\partial F}{\partial u}\right|_{u=\tilde{\omega}^{-}/k} & = & -\frac{(\omega/k-v/2)}{c_{i}^{3}\sqrt{2\pi}}\exp\left[-\frac{(\omega/k-v/2)^{2}}{2c_{i}^{2}}\right],
\end{eqnarray}
so that the growth rate of the instability can be approximated as
\begin{eqnarray}
\textrm{Im}\left(\frac{\omega}{\omega_{pi}}\right) & \simeq & \frac{1}{4}\sqrt{\alpha^{3}\pi}\left(\frac{\omega_{pi}}{kc_{i}}\right)^{3}\tilde{\omega}_{pi}\left\{ \left(\frac{kv}{2\omega_{pi}}-1\right)\exp\left[-\frac{(\omega_{pi}-kv/2)^{2}}{2k^{2}c_{i}^{2}}\right]\right.\nonumber \\
 &  & \left.-\,\left(\frac{kv}{2\omega_{pi}}+1\right)\exp\left[-\frac{(\omega_{pi}+kv/2)^{2}}{2k^{2}c_{i}^{2}}\right]\right\} ,
\end{eqnarray}
where we have set $\omega\sim\omega_{pi}$ on the right hand side.


\end{document}